\def\etal{{\it et al.\thinspace}}
\def\mearth{{\rm\,M_\oplus}}
\def\msun{{\rm\,M_\odot}}

\documentclass[apjl]{emulateapj}

\begin{document}

\shorttitle {Few habitable planets around low-mass stars} 
\shortauthors{Raymond, Scalo \& Meadows}

\title{A decreased probability of habitable planet formation  around
low-mass stars}

\author{Sean N. Raymond\altaffilmark{1,2,*}, John Scalo\altaffilmark{3,*},
\& Victoria S. Meadows\altaffilmark{4,*}}

\altaffiltext{1}{Center for Astrophysics and Space Astronomy, University of
Colorado, 389 UCB, Boulder CO 80309-0389; raymond@lasp.colorado.edu}
\altaffiltext{2}{NASA Postdoctoral Program Fellow.}
\altaffiltext{3}{Department of Astronomy, University of Texas, Austin, TX; parrot@as.utexas.edu}
\altaffiltext{4}{Infrared Processing and Analysis Center, California
Institute of Technology, Pasadena, CA; vsm@ipac.caltech.edu}
\altaffiltext{*}{Member of NASA Astrobiology Institute}

\begin{abstract}

Smaller terrestrial planets ($\lesssim 0.3\mearth$) are less likely to retain
the substantial atmospheres and ongoing tectonic activity probably required to
support life.  A key element in determining if suf{f}iciently massive
``sustainably habitable'' planets can form is the availability of solid
planet-forming material. We use dynamical simulations of terrestrial planet
formation from planetary embryos and simple scaling arguments to explore the
implications of correlations between terrestrial planet mass, disk mass, and
the mass of the parent star. We assume that the protoplanetary disk mass
scales with stellar mass as $M_{disk}\propto f\,M_\star^h$, where $f$ measures
the relative disk mass, and $1/2 < h < 2$, so that disk mass decreases with
decreasing stellar mass.  We consider systems without Jovian planets, based on
current models and observations for M stars.  We assume the mass of a planet
formed in some annulus of a disk with given parameters is proportional to the
disk mass in that annulus, and show with a suite of simulations of late-stage
accretion that the adopted prescription is surprisingly accurate.  Our results
suggest that the fraction of systems with suf{f}icient disk mass to form $>
0.3\mearth$ habitable planets decreases for low-mass stars for every realistic
combination of parameters.  This ``habitable fraction'' is small for stellar
masses below a mass in the interval 0.5 to 0.8 $\msun$, depending on disk
parameters, an interval that excludes most M stars.  Radial mixing and
therefore water delivery are inef{f}icient in lower-mass disks commonly found
around low-mass stars, such that terrestrial planets in the habitable zones of
most low-mass stars are likely to be small and dry.
\end{abstract}


\section{Introduction}

The circumstellar habitable zone (HZ) is the distance annulus within which a
terrestrial planet can maintain liquid water on its surface.  The average
distance of the HZ from the parent star depends primarily on the star's
luminosity, and is closer-in and narrower around lower-mass stars (Kasting
\etal 1993). For a planet to remain habitable, it must control its surface
temperature over long timescales, possibly via the CO$_2$-carbonate cycle
(Walker \etal 1981), which is enabled by a relatively thick atmosphere, a
hydrological cycle, and active plate tectonics. These planetary
characteristics are less likely to be sustained over long periods by less
massive terrestrial planets, or those with low water contents.  Whether
potentially habitable planets of suf{f}icient mass will form depends on the
surface density of rocky material in their parent protoplanetary disks. Disks
with higher surface densities tend to form a smaller number of more massive
planets (Wetherill 1996; Raymond \etal 2004, Kokubo \etal 2006), and can
increase terrestrial planet water content by scattering more distant,
water-rich material into the HZ (Morbidelli \etal 2000; Raymond \etal 2004,
2005a).

Here we investigate the formation, masses and water contents of terrestrial
planets as a function of stellar mass by assuming a parameterized relationship
between disk and stellar mass.  We adopt a simple model relating the HZ planet
mass to the disk mass, based on results in the literature.  We use a set of
numerical simulations of terrestrial planet growth around stars between 0.2
and 1 $\msun$ to test the validity of the model.  The simulations agree well
with the model, and justify an exploration of the consequences of such a
planet-disk-stellar mass relation, and in addition allow us to consider how
water delivery is af{f}ected as a function of stellar mass.

Our main result is to derive probabilistic limits on the stellar mass below
which sustainably habitable planets cannot form because of insuf{f}icient disk
mass available for accretion to a critical planet mass of 0.3 $\mearth$.  In
addition, we show how this stellar mass limit depends on the parameters of the
disk-star mass relationship, which is currently poorly constrained.  In a
statistical framework, we show that the fraction of stars capable of forming
habitable planets {\it in situ} decreases quickly with decreasing stellar
mass, and that terrestrial planets in the habitable zones of M stars ($M_\star
\lesssim 0.5 \msun$; e.g., Reid \& Hawley 2000) are likely to be small and
dry.  An interesting test of our models comes from the recent detection of a
$\sim 5 \mearth$ planet close to the habitable zone of the M dwarf Gliese 581
(Udry \etal 2007).  The implications of this system for our formation models
are discussed in section 4.

To simplify our calculations, we assume there are no gas giant (Jovian)
planets present, and we test the ef{f}ects of this assumption in section 3.3.
Systems without giant planets may represent the majority of planetary systems
with habitable planets.  Available estimates of the fraction of solar-like
stars with massive planets (Tabachnick \& Tremaine 2002, Lineweaver \& Grether
2004, Beer \etal 2004; Fischer \& Valenti 2005) allow for 70-90\% of stars to
have planetary systems without Jovians, and an even larger fraction for
lower-mass stars (Endl \etal 2006; Butler \etal 2006).  In addition, Scholz
\etal (2006) point out that estimated submm and mm outer disk masses show that
only a small fraction of low-mass stars or substellar objects have suf{f}icient
disk mass to form a planet with the mass of Jupiter (but see Hartmann \etal
2006).  The lack of a correlation between the presence of debris disks and the
metallicity of the parent star (Greaves \etal 2006; Beichman \etal 2006;
Moro-Mart\'in \etal 2007), unlike the correlation known for exogiant planets
(Gonzalez 1997, Santos \etal, Fischer \& Valenti 2005), also supports the
likelihood of gas-giant-free planetary systems. There are observational (Endl
\etal 2006; Butler \etal 2006) and theoretical (Ward \& Hahn 1995; Adams \etal
2004; Laughlin \etal 2004, Ida \& Lin 2005) reasons to think that the fraction
of M stars with Jovian planets is significantly smaller than for solar-mass
stars.

\section{Methods}

\subsection{Habitable Planet Mass Limit}
Following Williams \etal (1997), we assume a lower mass limit for planetary
habitability that supports active plate tectonics for several Gyr.
Significant mass is a necessary, but not suf{f}icient, requirement for
sustainable planetary habitability (e.g., Lissauer 1999).  Using the
radioactive flux limit from Williams \etal (1997), the critical planetary mass
scales like $\rho^{-2} e^{3 \lambda t}$, where $\rho$ is the planet's bulk
density, $\lambda=1.5 \times 10^{10} yr^{-1}$ for $^{238}$U, and $t$ is the
duration of tectonic activity. For an active lifetime of 5 Gyr, and iterating
on the density to be consistent with the resulting critical mass, we get a
critical mass of 0.3 $\mearth$ for a density of 4.5 $g\, cm^{-3}$.  This
result is sensitive to the plate tectonic activity timescale associated with
sustained habitability, which is fairly arbitrary.  The average age of stars
in the Galactic disk is about 5 Gyr, so the choice of 5 Gyr allows about half
of such stars with sustainably habitable planets to be on the main sequence
today, for a roughly constant star formation rate.  Throughout the paper we
consider 0.3 $\mearth$ as our critical mass limit for planetary habitability,
and we discuss the implications of changing this limit in section 4.

\subsection{Protoplanetary Disk Properties}

We assume the surface density of protoplanetary disks, $\Sigma$, to scale
with heliocentric distance $r$ as:
\begin{equation}
\Sigma(r) = \Sigma_1\, f\,Z \left(\frac{r}{\rm 1 AU}\right)^{-\alpha} \left(\frac{M_{\star}}{M_{\odot}}\right)^h,
\end{equation}
where $\Sigma_1$ is the surface density of solids at 1 AU in the minimum-mass
solar nebula (MMSN) model ($\approx 6 g \, cm^{-2}$), $h$ adjusts the density
scaling with $M_{\star}$, $f$ is a scaling factor for the disk ($f$=1 for
MMSN), $Z$ is the stellar metallicity (1.0=solar), and $\alpha$ moderates the
steepness of the density profile ($\alpha = 3/2$ in the MMSN model; Hayashi
1981, but more recent reconstructions using an accretion disk model as a
constraint gives $\alpha = 1/2$; Davis 2005).  Most disk models that include
viscous and irradiation heating give $\alpha$ between about 1/2 and 1 for the
inner disk (e.g. D'Alessio \etal 1998; Garaud \& Lin 2007, Dullemond et
al. 2007), and high-resolution submillimeter observations suggest $\alpha$
between 1/2 and 1 in the outer disk (Mundy \etal 2000, Looney et al. 2003,
Andrews and Williams 2006).  We take $\alpha = 1$ as a fiducial value but
consider values in the range 1/2 to 3/2.

Present estimates of disk masses in nearby star-forming regions (see Andre \&
Montmerle 1994, Eisner \& Carpenter 2003, Andrews \& Williams 2005) give a
large spread, with median or average masses usually somewhat below the MMSN
value, but a significant fraction more massive, including a very massive disks
observed in both Orion and Taurus (e.g. Williams \etal 2005, Eisner \&
Carpenter 2006).  At the other end of the mass spectrum, the estimates by
Williams \etal (2005) of mean masses for 18 Orion proplyds below the
$3-\sigma$ detection limit yielded $8 \times 10^{-4} \msun$, or only $3
\mearth$ in solids for a gas-to-dust ratio of 100.  There also exists a factor
of 3-5 range in metallicity for thin-disk stars (Nordstrom \etal 2004), which
adjusts the ef{f}ective gas-to-dust ratio. 

There is considerable uncertainty in the scaling between stellar mass
$M_{\star}$ and disk mass $M_{disk}$.  The most promising lead comes from
estimates of outer disk masses using submm and mm techniques (Beckwith et
al. 1990), which remain uncertain because of adopted dust temperatures,
opacities, distances, and radiative transfer model.  Scholz \etal (2006)
estimated disk masses for brown dwarf disks and combined their results with
previous determinations for stellar and substellar objects from 0.02 to 3
$M_\odot$.  Their Fig. 3 shows that the ratio of disk mass to stellar mass is
roughly constant with stellar mass, with a large amount of scatter.  This
hints that $h \approx 1$, with a large range in $f$.  Many of the data points
from Scholz \etal (2006) are only upper limits, but we think extreme cases
$h=0$ or $h>2$ would be apparent in this data.

Additional clues come from the observed stellar accretion rate $\dot{M}$,
which scales as $\dot{M}\propto M_{\star}^2$, with a large dispersion
(Muzerolle \etal 2003, 2005, Natta \etal 2006).  It might therefore seem
natural to assume that $M_{disk} \propto M_\star^2$, i.e. $h$=2, assuming that
the viscous timescale is independent of stellar mass (e.g., Ida \& Lin 2005).
On the contrary, nearly all accretion disk models predict $\dot{M}\propto
M_{\star}^a$ with $a \sim 1$.  This matter is still unresolved; several
solutions have been proposed (see Padoan \etal 2006; Alexander and Armitage
2006; Dullemond \etal 2006; Hartmann \etal 2006; Gregory \etal 2007) including
selection/detection limitations (Clarke \& Pringle 2006).  Most of these
models are consistent with $h \approx 1$.  Although $h$ is currently poorly
constrained, we emphasize its importance for understanding whether sustainable
HZ planets can form around low-mass stars.  Given the results of Scholz \etal
(2006), we choose $h=1$ as our fiducial case, but explore the ef{f}ects of $h$
from 0.5 to 2.  


\subsection{Simulations and Models}

We performed dynamical simulations to evaluate the scaling between terrestrial
planet mass and disk properties.  For all runs, our starting conditions
reflect our fiducial case, with $fZ=1$, $h=1$, and $\alpha=1$.  We vary the
mass of the central star from 0.2 to 1 $\msun$.  Given the surface density
profile from Eq (1), we assume that planetary embryos have formed throughout
the inner disk of each star.  We assume that the disk evolved following the
standard model of terrestrial planet growth (e.g. Chambers 2004): grains
coalesced to form km-sized planetesimals; and embryos formed from
planetesimals via runaway and oligarchic growth (e.g., Kokubo \& Ida 1998),
spaced by 5-10 mutual Hill radii ($R_{H,m} = 0.5 (a_1 + a_2) (M_1+M_2/3
M_\star)^{1/3}$; $a_1$ and $a_2$ are the orbital radii and $M_1$ and $M_2$ the
masses of two adjacent embryos).  For the case of 1 $\msun$, we generated a
population of $\sim$ 75 embryos from 0.5 to 4 AU, totaling 4.95
$\mearth$.\footnote{Note that the outer edge of 4 AU for the embryo disk is
chosen based on previous simulations showing that the feeding zone of a
terrestrial planet in the habitable zone does not extend beyond 4 AU (Raymond
\etal 2007).}  For other stellar masses, we assume the population of embryos
has the same temperature, as we are interested in HZ planets (defined to have
the same temperature), and the feeding zone boundaries should scale in the
same way as the HZ mean distance.  We therefore scale the inner and outer
boundaries by the stellar flux (i.e., as the luminosity $L_\star^{1/2}$),
using a mass-luminosity relation from Scalo \etal (2007) that is a fit to data
from Hillenbrand \& White (2004): $y = 4.101 x^3 + 8.162 x^2 + 7.108 x +
0.065$, where $y=log(L_\star)$ and $x=log(M_\star)$.  Table 1 summarizes our
starting conditions for each stellar mass.  Note the variations in the number
of embryos $N$ included in each simulation, from $\sim$ 75 for the 1 $\msun$
simulations to almost 200 for the 0.2 $\msun$ simulations.  This variation in
$N$ is due to the Hill radius being very small at the small orbital distances
studied for lower-mass stars.  In other words, models of embryo growth predict
that there really are more embryos in these inner disks than in the more
distant regions studied for higher-mass stars (e.g., Kokubo \& Ida 1998).
Kokubo \etal (2006) showed that the bulk properties of accreted planets are
not particularly sensitive to such variations in $N$, so we are not concerned
that our choice of embryo formation models will af{f}ect our results.  Note also
that the $0.2 \msun$ star has less than 0.3 $\mearth$ in its entire disk, so
it serves primarily to tell us about the ef{f}iciency with which the disk mass
is used to make planets under these conditions, but cannot form a habitable
planet according to our criterion.  Each numerical experiment was run
independently four times starting from dif{f}erent random initial embryo
positions for a total of twenty simulations.  We give embryos random starting
eccentricities $<$0.02 and inclinations $<$0.1 degrees.

Embryos are assigned water contents based on values from our Solar System,
where asteroids beyond $\sim$2.5 AU contain significant quantities of water
(Abe \etal 2000; see Fig. 2 in Raymond \etal 2004).  We assumed that this
boundary between ``dry'' and ``wet'' embryos (called the ``water line'' in
Table 1) scales with the stellar flux.\footnote{We use the term ``water line''
instead of the more common ``snow line'' because we are not necessarily
assuming this to be the location where the temperature drops below $\sim 170$
K and water ice can condense (2.7 AU in the MMSN; Hayashi 1981).  Rather, we
are simply assuming this boundary to divide dry and wet material.  The water
line may actually be located somewhat interior to the snow line, as water-rich
bodies can be shifted inward via either gas drag or by eccentricity pumping
during embryo formation (e.g., Cyr \etal 1998; Kokubo \& Ida 1998).}  Embryos
beyond this boundary contain 5\% water by mass, and those inside are dry
(Table 1).  We integrated each simulation for 200 Myr using the hybrid
integrator {\it Mercury} (Chambers 1999).  Timesteps were chosen to sample the
innermost body's orbit at least twenty times (e.g., Rauch \& Holman 1999;
Levison \& Duncan 2000; see Table 1), and so varied with each set of
simulations, from 6 days for 1 $\msun$ to 0.2 days for 0.2 $\msun$.
Collisions were treated as inelastic mergers conserving water.

\section{Results}

\subsection{Terrestrial Planet Mass vs Stellar Mass}

\begin{figure}
\centerline{\epsscale{0.7}\plotone{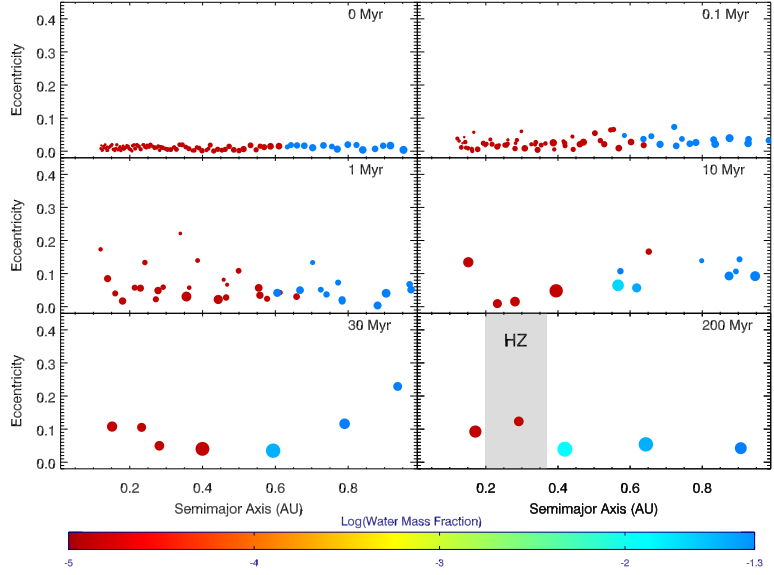}}
\caption{Evolution of a simulation around a 0.6 $\msun$ star.  Particle size
scales with the planetary size as the planet mass$^{1/3}$, but is not to scale
on the x axis.  Colors correspond to water contents, from red (dry) to blue
(5\% water by mass -- see color bar).  For scale, the planet that formed at
0.29 AU in the HZ is 0.06 $\mearth$ and the planet at 0.42 AU is 0.21
$\mearth$.  The HZ is shaded in the final panel of the simulation. Note that,
although the HZ planet is dry, some water delivery did occur, as two
water-rich embryos were accreted by the planet at 0.41 AU.}
\label{fig:aet}
\end{figure}

Figure~\ref{fig:aet} shows the evolution of a simulation for a $0.6 \msun$
star: the disk is excited by gravitational perturbations among the embryos.
As eccentricities increase, orbits cross and collisions occur.  In time, a few
planets grow and the number of bodies dwindles.  Embryos may be scattered far
from their original locations, sometimes delivering water-rich material to
planets in the inner regions.  Water delivery occurs relatively late, because
multiple scattering events are needed for significant radial movement (Raymond
\etal 2007).  In this case some water delivery occurred: a 0.21 $\mearth$
planet formed at 0.41 AU, just beyond the outer boundary of the HZ, accreted
two water-rich embryos originating beyond the water line at 0.61 AU.  However,
no water was delivered to any planets in the HZ in this case.  This is a
typical outcome for the low-mass disks expected to be common around low-mass
stars (see below).

Figure~\ref{fig:mall} shows the final outcome of ten simulations, with the
Solar System included for scale (the Earth's water content is $\sim 10^{-3} by
mass$; L\'ecuyer \etal 1998).  It is clear that, for our assumptions of $h =
1$, $\alpha = 1$ and $h =1$, terrestrial planets are much smaller around
low-mass stars.  In addition, planets that form in the HZs of low-mass stars
tend to be dry, and more closely spaced.  Note that, although we do not follow
their orbits for a full 5 Gyr, we assume water-rich planets that form in the
HZ to be potentially habitable.  Very late-stage instabilities were not seen
in any of these simulations, nor in the $>$Gyr integrations from Raymond \etal
(2005a, 2006a), but the potential disruption of the system at times as late as
5 Gyr cannot be ruled out.  In addition, a late-stage instability could
potentially alter the orbit of a distant planet, causing it to collide with a
planet in the HZ.  Such an event could even deliver water at a very late
stage.  Although this is certainly possible, we have found this type of event
to be rare in previous long-term simulations.  In addition, the source of such
a rogue planet would have to be quite far out, because accretion tends to
occur faster closer to the star, and crossing orbits are needed to cause a
strong scattering event.

Figure~\ref{fig:mpl} shows the mean mass of simulated planets that formed in
the HZ as a function of stellar mass, with error bars representing the range
of outcomes from the four simulations for a given stellar mass $M_\star$.  The
solid curve shows the prediction of a simple scenario in which the mass of a
planet in the HZ is proportional to the mass contained within the HZ annulus,
such that
\begin{equation}
M_{pl} \propto \frac{\Sigma_1 f Z
M_\star^h}{2-\alpha}\left(r_{HZ,out}^{2-\alpha} - r_{HZ,in}^{2-\alpha}\right),
\end{equation}
\noindent where $r_{HZ,in}$ and $r_{HZ,out}$ are the inner and outer
boundaries of the HZ (see Table 1).  This model is a very simple, but not
unreasonable, approximation for planet mass, and the quality of the fit is
remarkable -- note that this model was calibrated such that a disk with $fZ=1$
will form a 1 $\mearth$ planet in the HZ around a solar-mass star.  Kokubo
\etal (2006) showed that the planet mass scales roughly linearly with the
available mass, albeit for a fixed stellar mass, and we independently found
the same result for $M_\star = 0.4 \msun$.  In fact, Kokubo \etal (2006) found
a slightly stronger than linear correlation, $M_{pl} \propto
M_{disk}^{0.97-1.1}$.  Fig~\ref{fig:mall} shows that the mean interplanetary
spacing decreases somewhat for lower-mass stars, and that there are slightly
more planets for the lower-mass stars. We know that the total mass in the
habitable zone, $M_{HZ}$, is equal to the number of planets, $N$, times the
average planet mass, $M_{pl}$.  Our simulations suggest that $N \propto
M_{HZ}^{-0.1}$ or so.  Indeed, a model with $M_{pl} \propto M_{HZ}^{1.1}$
provides a fit that is comparable to the one in Fig.~\ref{fig:mall}.  However,
for the remainder of the paper we assume that $M_{pl} \propto M_{HZ}$ (Eqn 2).
The reason for this assumption has to do with the goals of the paper.  We are
attempting to constrain the locations in $M_\star - h - fZ$ parameter space
that might harbor potentially habitable planets with $M_{pl} \geq 0.3
\mearth$.  To be conservative in our evaluations, we prefer to slightly
overestimate, rather than underestimate, $M_{pl}$.  The dif{f}erence between
the two estimates is negligible for larger stellar masses, but is as much as
$\sim$ 40\% below 0.1 $\msun$.

The planet mass decreases monotonically with stellar mass for all reasonable
parameter values (Eq. 2; Fig~\ref{fig:mpl}), with a scatter in the details of
a given system based on the stochastic nature of the accretion process (e.g.,
Wetherill 1996).  Only for very steep density profiles ($\alpha > 2$) or
reversed disk mass scalings ($h<0$) can the planet mass increase at lower
stellar masses.  These ef{f}ects are the result of the strong dependence of
the HZ's location's on stellar luminosity, and therefore on stellar mass.  For
a given value of $f$, $Z$, $\alpha$, and $h$, there exists a stellar mass
limit below which the formation of a $>0.3 \mearth$ planet in the HZ is
unlikely.  For $\alpha = 1$ and $h=1$, this limit ranges from 1 $\msun$ for
$fZ < 0.3$ to 0.74 $M_\odot$ for $fZ=1$ to 0.43 $M_\odot$ for $fZ=5$.  These
limits clearly depend on the critical mass for habitability; for instance, the
limit is 0.53 $\msun$ for the $fZ=1$ case if the critical habitable mass is
0.1 $\mearth$.  Recall that $fZ$ represents a scaling of the disk mass, i.e.,
the disk's relative mass $f$ times the relative abundance of solids, assumed
to scale with the stellar metallicity $Z$.

Figure~\ref{fig:mstar-h} shows the location in $M_\star-fZ$ space where
planets $>0.3 \mearth$ can form in the HZ, assuming $\alpha=1$.  Each curve
corresponds to a given value of $h$; planets $> 0.3 \mearth$ form above and to
the right of each curve.  More massive or metal-rich disks can form habitable
planets around lower-mass stars.  In addition, it is easier to form $> 0.3
\mearth$ planets in the HZ for more centrally-condensed disks, i.e., for
larger values of $\alpha$ (not shown in Fig.~\ref{fig:mstar-h}).  Given the
large amount of variation in $fZ$ and other uncertainties, we do not consider
these limits to be firm.  However, given the large uncertainties and expected
variation in $f$ and other parameters, we do not consider these limits to be
meaningful except in a statistical sense.  For an ensemble of disks, the
fraction of $>0.3 \mearth$ planets that form decreases significantly for
low-mass stars.  A probabilistic version of the mass limit estimate is
discussed further in \S 4.


\subsection{Formation Timescales and Planetary Water Contents}

Figure~\ref{fig:tf} shows the mean formation timescales for HZ planets in our
simulations.  Around 0.2 $\msun$ stars, terrestrial planets in the HZ form in
a few Myr.  This increases to 20-50 Myr for Sun-like stars, consistent with
estimates from Hf/W isotopic measurements (e.g., Jacobsen 2005).  The reason
for the speedup in accretion times in the HZs of low-mass stars is due to a
combination of the faster orbital speeds in the HZs of low-mass stars and the
higher surface densities (albeit much lower total HZ masses).  For instance,
if we assume $L_\star \propto M_\star^4$ (a rough fit to Hillenbrand \& White
2004), then the location of the HZ, $r_{HZ}$, scales with the stellar mass as
$r_{HZ} \propto L_\star^{1/2} \propto M_\star^2$.  The orbital speed in the
HZ, $v_{HZ}$, scales as $v_{HZ} \propto (M_\star/r_{HZ})^{1/2} \propto
M_\star^{-1/2}$.  The surface density in the HZ, $\Sigma_{HZ}$, scales as
$\Sigma_{HZ} \propto r_{HZ}^{-\alpha} M_\star^h \propto M_\star^{-2\alpha+h}
\propto M_\star^{-1}$ for our case of $\alpha = h = 1$.  So, using a very
rough approximation that the growth time $t_G$ scales inversely with the
product of the orbital frequency ($\sim v_{HZ}/r_{HZ}$) and the local surface
density (Safronov 1969) yields $t_G \propto M_\star^{7/2}$ for our case.  This
scaling is a very poor match to our simulations, yielding accretion timescales
that are far too short for stellar masses below $0.6 \msun$ (dotted line in
Fig.~\ref{fig:tf}).  This is because the final stage of planetary growth is
dominated by isolated scattering events between embryos rather than accretion
from a continuous medium of planetesimals.  In fact, we notice that a slightly
dif{f}erent scaling, with $t_G \propto (\Sigma_{HZ} v_{HZ})^{-1} \propto
M_\star^{3/2}$, provides a good empirical fit to our simulated formation times
(dashed line in Fig.~\ref{fig:tf}).  We suspect that this is simply because of
the dynamics of this accretion regime, in which the relevant bodies are not on
initially crossing orbits, and secular perturbations are required to excite
eccentricities and thus to cause orbits to cross.

Lissauer (2007) argued that the formation time of HZ planets around low-mass
stars is very short, a few hundred thousand years for a 1/3 $\msun$ star.  Our
simulations confirm the trend that HZ planets form faster around low-mass
stars.  However, Fig.~\ref{fig:tf} shows that these formation times do not
scale nearly as strongly with stellar mass as indicated by Lissauer (he
calculated $t_G \sim M_\star^{6.2}$).  The reason for this discrepancy is that
Lissauer required an Earth-mass planet to form in the HZ, so to accomplish
this he increased the disk mass by a large factor.  Our models suggest that a
factor of 20-30 increase is needed to form a 1 $\mearth$ planet in the HZ of a
$1/3 \msun$ star; as expected, that factor corresponds to the approximate
dif{f}erence in accretion times between our simulations and Lissauer's.
However, the fraction of 1/3 $\msun$ stars with disks massive enough to form a
1 $\mearth$ planet in the HZ is less than 5\%, although that fraction
increases to about 15\% if $\alpha=3/2$ (see Fig.~\ref{fig:frac} below).  In
addition, recall that the much shallower empirical scaling ($t_G \propto
M_\star^{3/2}$ for our case) provides a far better fit to our simulations than
the simple derivation using the orbital frequency ($t_G \propto M_\star^{7/2}$
for our case; based on Safronov 1969 and calculated in similar fashion to
Lissauer 2007).

The mean water content of our HZ planets decreases drastically around low-mass
stars.  In fact, only one out of 19 HZ planets formed from a star with
$M_\star < 0.6 M_\odot$ contained a significant amount of water, compared with
two out of six HZ planets for $M_\star = 0.8 \msun$ and two out of four for
$M_\star = 1 \msun$ (see Fig.~\ref{fig:mall}). The reason for this trend --
drier planets in the HZs of lower-mass stars -- is inef{f}icient dynamical
stirring in the low-mass disks found around low-mass stars, because more
massive embryos are needed to increase eccentricities enough for significant
radial mixing and therefore water delivery (Morbidelli \etal 2000; Raymond
\etal 2004, 2007).  Lissauer (2007) argued that planets in the HZs of M stars
would be deficient in volatiles for three reasons: 1) collision velocities are
higher; 2) formation timescales are faster; and 3) pre-main sequence evolution
of the lowest-mass M stars is slow, such that the snow line and HZ move inward
on a Myr to Gyr timescale (see Kennedy \etal 2006).  We agree with point 1:
collision velocities are proportional to the orbital speed, so the velocity of
impactors at infinity (neglecting the escape speed) scales with stellar mass
roughly as $M_\star^{-1/2}$ (see above).  We also agree qualitatively with
point 2: formation timescales are indeed likely to be shorter around low-mass
stars by perhaps an order of magnitude compared with solar-mass stars
(Fig.~\ref{fig:tf}).  Point 3 is more uncertain: the inward movement of the
snow line could af{f}ect the water contents of HZ planets if the formation
timescale is shorter than the snow line's drift timescale.  In such a case,
material in a given region may be too hot to contain water at early times,
during accretion.  After several Myr, that zone may drop below the threshold
for water to condense.  If, however, the planets are fully formed by this
time, then water delivery is impossible.  Note that this pre-main sequence
scenario applies only to very low-mass stars.  Our results suggest that this
is a moot point, as HZ planets around such very low-mass stars are unlikely to
have wide enough feeding zones to accrete water-bearing material, even at late
times.

Thus, we argue that terrestrial planets in the HZs of low-mass stars are
likely to be dry, but for a dif{f}erent reason than Lissauer.  Simply put, very
little water-rich material is likely to impact such planets at all, because
radial mixing is inef{f}icient in low-mass disks preferentially found around
low-mass stars.  The influence of additional gas or ice giant planets in the
system is not expected to increase the water contents of HZ planets, at least
not from the asteroidal source of water considered here (S. Raymond 2007, in
preparation).  However, migrating giant planets (not modeled here) can stir up
eccentricities and induce the formation of very water-rich planets in their
wake (Raymond \etal 2006b; Fogg \& Nelson 2007; Mandell \etal 2007).  In
addition, subsequent delivery of water from a cometary source is possible.

\subsection{Ef{f}ects of Giant Planets}

An obvious criticism of this work is the lack of giant planets.  Simulations
have shown that giant planets play an important role in terrestrial planet
formation (e.g., Wetherill 1996; Levison \& Agnor 2003).  Giant planets on
orbits exterior to the terrestrial region, such as Jupiter and Saturn, stir up
the eccentricities of embryos from the outside in via secular and resonant
perturbations, while mutual scattering of embryos excites eccentricities from
the inside out (e.g., Raymond \etal 2005b).  Accretion is suppressed in the
vicinity of the giant planet, because perturbations between embryos during
their growth can scatter them into unstable regions such as strong giant
planet resonances (Wetherill 1994).  Indeed, the combined ef{f}ects of embryo
scattering and perturbations from Jupiter and Saturn are thought to be the
cause of the depletion of the Solar System's asteroid belt (Wetherill 1992;
Chambers \& Wetherill 2001; Petit \etal 2001).

The net ef{f}ect of giant planets is to increase the eccentricities of embryos
during accretion.  This can cause the width of the feeding zone of terrestrial
planets to be increased, causing planets to be somewhat more massive and less
numerous than in the absence of giant planets (Levison \& Agnor 2003).  This
increase in planet mass can be as high as $\sim 30\%$ (or planet masses can be
decreased), but occurs primarily for relatively weak perturbations, i.e., for
less massive giant planets on relatively circular orbits (S. Raymond 2007, in
preparation).  Even on circular orbits, external giant planets rarely enhance
water delivery.  Rather, their perturbations clear out external, water-rich
material leaving behind relatively dry terrestrial planets (S. Raymond 2007,
in preparation).  Indeed, terrestrial planets in systems with giant planets on
eccentric orbits are likely to be dry (Chambers \& Cassen 2002; Raymond \etal
2004; Raymond 2006).  Note that systems with close-in giant planets may also
contain Earth-like planets, which should be very water-rich (Raymond \etal
2006b; Mandell \etal 2007).

To test the ef{f}ects of giant planets, we performed eight additional
simulations including one giant planet on an exterior orbit.  We chose the 0.6
$\msun$ case for these simulations, to see if giant planet perturbations could
increase the mean terrestrial planet mass above $0.3 \mearth$.  We used the
same four sets of embryos described in Table 1 and included a giant planet on
a circular orbit at 1.3 AU, corresponding roughly to the orbital distance with
the same temperature (and therefore at a comparable dynamical separation from
the HZ) as Jupiter for a 0.6 $\msun$ star.  We ran two sets of simulations,
one each for a Neptune- and a Jupiter-mass giant planet.

Planets that formed in the HZ of our 0.6 $\msun$ simulations without giant
planets had masses between 0.06 and 0.18 $\mearth$, with a mean of 0.10
$\mearth$.  Systems with a Jupiter- [or Neptune-] mass giant planet formed,
respectively, HZ planets with masses between 0.05 [0.06] and 0.22 [0.10]
$\mearth$, with a mean of 0.11 [0.09] $\mearth$, very close to the no giant
planets cases.  None of the seven HZ planets in systems with no giant planets
contained any water-rich material.  Two out of eleven HZ planets that formed
in the eight giant planet systems contained a substantial amount of water.
While this is only a small fraction of outcomes, it does show that giant
planet stirring can, in some cases, help in water delivery.  However, in the
vast majority of situations, giant planets either hinder or have no effect on
water delivery (S. Raymond 2007, in preparation).  The formation timescales of
HZ planets in giant planet simulations were comparable to, or even slightly
longer than, those with no giant planets.  In almost all cases, planets with
or without giant planets reached 75\% of their final masses within 10-20 Myr
(see Fig.~\ref{fig:tf}).  We suspect that the reason giant planets do not
accelerate accretion in this case is because more radial mixing is happening,
such that accretion is somewhat less confined to a given annular zone.  This
ef{f}ect appears to be small, as less than 20\% of planets in the giant planet
simulations contain water.

\section{Discussion and Conclusions}

Our analysis is based on three key assumptions: 1) terrestrial planets
below a given mass are unlikely to sustain life on Gyr timescales -- following
Williams \etal (1998), we derive a limiting planet mass of roughly $0.3
\mearth$; 2) a protoplanetary disk can be well described by just a few
parameters (see Eq. 1): the radial surface density slope $\alpha$, the
relative disk mass in solids $fZ$, and the exponent characterizing the disk
mass-stellar mass relationship, $h$; and 3) the typical planet mass in the
habitable zone is proportional to the disk mass in that zone (see Eq. 2 and
Fig.~\ref{fig:mpl}).  We performed a suite of twenty dynamical simulations of
the late-stage growth of terrestrial planets that confirmed that this scaling
does hold for a wide range of stellar masses.  We assumed in our calculations
that no giant planets were present.  We tested that assumption for the 0.6
$\msun$ case and found that the ef{f}ects of exterior giant planets do not
increase the masses or decrease the formation times of terrestrial planets, at
least for the cases considered here.

Starting from these assumptions, we explored the range of parameters that
allow planets more massive than $0.3 \mearth$ to form in the habitable zones
(HZs) of their host stars.  We found that any realistic combination of
parameters led to a stellar mass - planet mass correlation such that HZ planet
masses decrease with decreasing stellar mass (see Fig.~\ref{fig:mpl}).  For a
given set of parameters ($\alpha$, $h$,$fZ$), one can derive a stellar mass
below which the probability of a habitable planet forming in a disk with those
parameters decreases to zero (see Fig.~\ref{fig:mstar-h}).  For instance, for
our fiducial parameters ($h=1$, $\alpha=1$,$f=1.2$, chosen so that the
simulations produce a mean planet mass of 1 $\mearth$ for a 1 $\msun$ star),
this critical stellar mass is $0.7 \msun$.  This would mean that no M stars
and few K stars should have sustainably habitable planets.  However, there are
some caveats to this approach.  First, for main sequence stars, it is
impossible to go back in time to learn the properties of their protoplanetary
disks, most importantly $f$ and $\alpha$.  Second, for a given set of
parameters, there still exists a range of potential outcomes because of the
stochastic nature of the final stages of planetary growth (e.g., Wetherill
1996, Raymond \etal 2004; see Fig.~\ref{fig:mall}).  Third and most
importantly, there is no universal set of parameters that applies for all
disks.  A given set of parameter values, corresponding to a specific disk or
set of disks, has a corresponding planet mass distribution.  For instance, the
above-mentioned 0.7 $\msun$ limit decreases to $0.31 \msun$ for a more massive
disk with $fZ = 10$, and would decrease farther still for $h<1$ or $\alpha
>1$.  Thus, our approach is only viable in a statistical sense, by combining
likely outcomes from a distribution of disks with varying properties.

What are the values of the crucial parameters?  We expect $h$ and $\alpha$ to
be universal values characterizing the distributions of all disks, but $f$ and
$Z$ to vary from disk to disk.  As discussed above, there is considerable
uncertainty in the value of $h$, which controls the disk mass-stellar mass
scaling.  If $h$ were larger than 2 or smaller than 0.5, we expect that the
trend would be apparent in Fig. 3 in Scholz \etal (2006).  Thus, a roughly
linear disk mass-stellar mass relationship ($h=1$) seems likely.  A range of
roughly 2 orders of magnitude in disk mass is observed around a given stellar
mass (e.g., Eisner \& Carpenter 2003; Williams \etal 2005; Scholz \etal 2006),
implying a corresponding range in $f$ values.  Note, however, that there
exists a large uncertainty in tying a given $f$ value to an absolute disk mass
or surface density, because of both observational and theoretical
uncertainties (e.g., Carpenter \etal 2005).  There exists a range of 3-5 in
metallicity $Z$ for thin-disk stars (Nordstrom \etal 2004), although the range
for currently-forming stars may be much smaller, only $\sim$20\% (see
Cartledge et al. 2006).  Nearly all disk models and observations suggest that
the surface density exponent $\alpha = 0.5-1.5$ (e.g., Dullemond \etal 2007;
Andrews \& Williams 2006).  Our choice of $\alpha =1$ as a fiducial case was a
compromise between these varying estimates.

Given the large amount of intrinsic variation in the disk mass and accretion
outcome, as well as the uncertainty in $h$ and $\alpha$, we do not consider
any stellar mass limits from Fig.~\ref{fig:mstar-h} to be firm.  Rather, our
primary result is a statistical observation that, in our framework, the
fraction of disks capable of forming planets of a given mass in the habitable
zone decreases around low-mass stars.  We investigate a model with $fZ$
distributed as a Gaussian in log space, with a mean of $fZ=1$ and a standard
deviation of 0.5 dex.  For simplicity, all other parameters are held fixed,
with $\alpha=1$, and $h$=1.  Figure~\ref{fig:frac} shows that the fraction of
systems that can form a $>0.3 \mearth$ HZ planet decreases from 80\% around
solar-type stars to 42\% around 0.7 $M_\odot$ stars to 6\% around 0.4
$M_\odot$ stars to only 0.3\% around 0.2 $M_\odot$ stars.  Disks with more
centrally concentrated mass distributions (larger $\alpha$ values) have less
dif{f}iculty forming massive planets in HZs very close to low-mass stars (e.g.,
Raymond \etal 2005b).  For our chosen fiducial parameters, we estimate that
only 25\% of systems with $M_\star = 0.6 \msun$ can form planets $> 0.3
\mearth$, and only about 5\% for $M_\star =0.4 \msun$.  Disks with radial
surface density profiles as flat as $r^{-1/2}$ have near zero probability for
stars with $M_\star < 0.5 \msun$, while probabilities are 2-4 times larger for
disks with $r^{-3/2}$ profiles.  Reducing the required planetary mass for
habitability to $0.1 \msun$ increases the probability by a factor of 2-4 (see
below).

Our estimate of the minimum planet mass that can sustain enough tectonic
activity for a period of time that we took as 5 Gyr is also very uncertain.
It is simple to show that if the mass luminosity relation is of the form
$L_\star \propto M_\star^p$, then the derived critical stellar mass
$M_{\star,cr}$ below which there should be few planets with masses above a
critical mass $M_{pl,cr}$ is given by $M_{\star,cr} \sim (M_{pl,cr}/f)^B$,
with $B=2/[p(2-\alpha) + 2h]$.  Thus, a reduction in the adopted critical mass
for sustained habitability from 0.3 to 0.2 $\mearth$ reduces the required
value of disk mass normalization by the same factor, in order to obtain the
same critical stellar mass.  Alternatively, for a given chosen disk mass
normalization, the derived $M_{\star,cr}$ varies with the assumed planet mass
to a power that is between about 0.3 to 0.7 for most of the range of
parameters described above. The inset of Fig.~\ref{fig:frac} shows the ef{f}ect
of changing our $0.3 \mearth$ lower limit, retaining fixed values of $h=1$ and
$\alpha = 1$. 

A system of three planets with minimum masses between 5 and 15 $\mearth$ was
recently discovered orbiting the 0.31 $\msun$ star Gliese 581 (Bonfils \etal
2005; Udry \etal 2007).  Figures~\ref{fig:mstar-h} and~\ref{fig:frac} suggest
that such a low-mass star is unlikely to form massive terrestrial planets.
How can we reconcile this?  We see two possible explanations: 1) the planets
of Gl 581 formed {\it in situ} from a very massive disk, with $f Z \sim
30-50$; or 2) our assumption of {\it in situ} formation failed in this case:
the planets of Gl 581 formed farther from the star (probably in the icy
regions of a relatively massive disk, with $f Z$ of at least 5-10) and
migrated inward (Goldreich \& Tremaine 1980).  If any of the Gl 581 planets
were to transit the parent star, then it would be possible to derive a rough
composition (rocky vs. icy: Valencia \etal 2007; Fortney \etal 2007; Sotin
\etal 2007), and to distinguish between these two formation scenarios (Gaidos
\etal 2007; S. Raymond \etal 2007, in preparation).  Note that, even if the Gl
581 planets formed {\it in situ}, we expect that they would contain a
substantial amount of water.  Following the arguments made above, massive
disks promote eccentricity growth and radial mixing such that planets in the
HZ should accrete a large amount of water-rich material.

Even if HZ planets of suf{f}icient mass can form around M stars, they face
other challenges, including loss of atmosphere by intense stellar activity
coupled with their small distance to their parent stars, depending on the
strength of the planetary magnetic field (Lammer \etal 2007), a snow line that
takes perhaps a Gyr to get close enough to the HZ to give suf{f}icient water
(Kennedy \etal 2006), or loss of volatiles due to larger impact speeds and
faster formation times (Lissauer 2007; but see \S 3.2 and
Fig.~\ref{fig:tf}). However these latter two points should not af{f}ect
decisions concerning M star planet searches, since they only apply to the very
lowest-mass stars, $\sim 0.1-0.2 \msun$, whose apparent brightness and HZ
angular separation are far too small for any planned search (Scalo \etal
2007).  Unless the masses of disks are larger than is currently thought by a
significant factor and/or the critical planet mass for habitability is
considerably smaller, only a small fraction of accessible M star systems (with
$M_\star \gtrsim 0.3-0.4 \msun$) should have habitable planets that remain
habitable for billions of years.

\section{Acknowledgments}

This paper benefited from the referee's thoughtful comments, and from
discussions with Jim Kasting.  This work was performed by the NASA
Astrobiology Institute's Virtual Planetary Laboratory Lead Team, supported via
NASA CAN-00-OSS-01. J.M.S. was supported by NASA Exobiology grant
NNG04GK43G. S.N.R. was supported by an appointment to the NASA Postdoctoral
Program at the University of Colorado's Center for Astrobiology, administered
by Oak Ridge Associated Universities through a contract with NASA.
Simulations were performed at Weber State University and JPL using CONDOR
(www.cs.wisc.edu/condor).


\scriptsize
\begin{deluxetable}{ccccccc}
\tablewidth{0pt}
\tablecaption{Initial Conditions for Simulations\tablenotemark{1}}
\renewcommand{\arraystretch}{.6}
\tablehead{
\colhead{$M_\star$} &  
\colhead{Range (AU)} & 
\colhead{Mass ($\mearth$)} &
\colhead{N} &
\colhead{Timestep (d)} &
\colhead{HZ (AU)} &
\colhead{Water line (AU)}}
\startdata
1.0 & 0.5-4 & 4.95  & 75 & 6 & 0.8-1.5 & 2.5 \\
0.8 & 0.25-2 & 1.98 & 75 & 2.5 & 0.39-0.74 & 1.23 \\
0.6 & 0.12-1 & 0.75 & 100 & 1.0 & 0.20-0.37 & 0.61 \\
0.4 & 0.06-1 & 0.53 & 130 & 0.4 & 0.10-0.19 & 0.32 \\
0.2 & 0.03-0.5 & 0.13 & 190 & 0.2 & 0.05-0.1 & 0.16 \\
\enddata
\tablenotetext{1}{Columns are labeled as follows: `$M_\star$' is the stellar
mass in Solar masses; `Range' is the initial radial distribution of embryos;
`Mass' represents the total mass in embryos; `N' is the number of embryos;
`Timestep' is the timestep used for the integration in days; `HZ' is the
extent of the habitable zone for each case; and `Water line' indicates the
boundary beyond which embryos are assumed to contain 5\% water by mass.}
\end{deluxetable}
\normalsize 


\begin{figure}
\centerline{\epsscale{0.7}\plotone{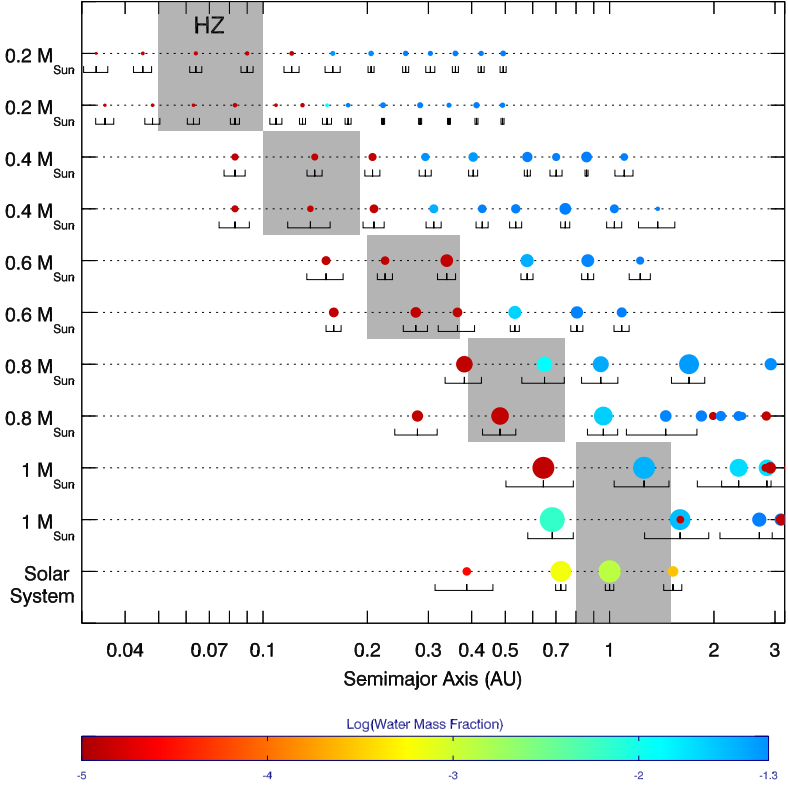}}
\caption{Final outcomes of ten simulations, two for each stellar mass chosen,
and with the Solar System for scale. As in Fig~\ref{fig:aet}, color represents
water content, and particle size scales with the planet size, i.e., the planet
mass$^{1/3}$.  The lines under each planet represent the radial excursion of a
planet over its orbit, i.e. its orbital eccentricity -- these values are
averaged over the last 10 Myr of each simulation.  Solar System water contents
are from Lodders \& Fegley (1998) and L\'ecuyer \etal (1998), and
eccentricities are 3 Myr averages from Quinn \etal (1991).  The HZ is shaded
for each stellar mass. }
\label{fig:mall}
\end{figure}

\begin{figure}
\centerline{\epsscale{0.9}\plotone{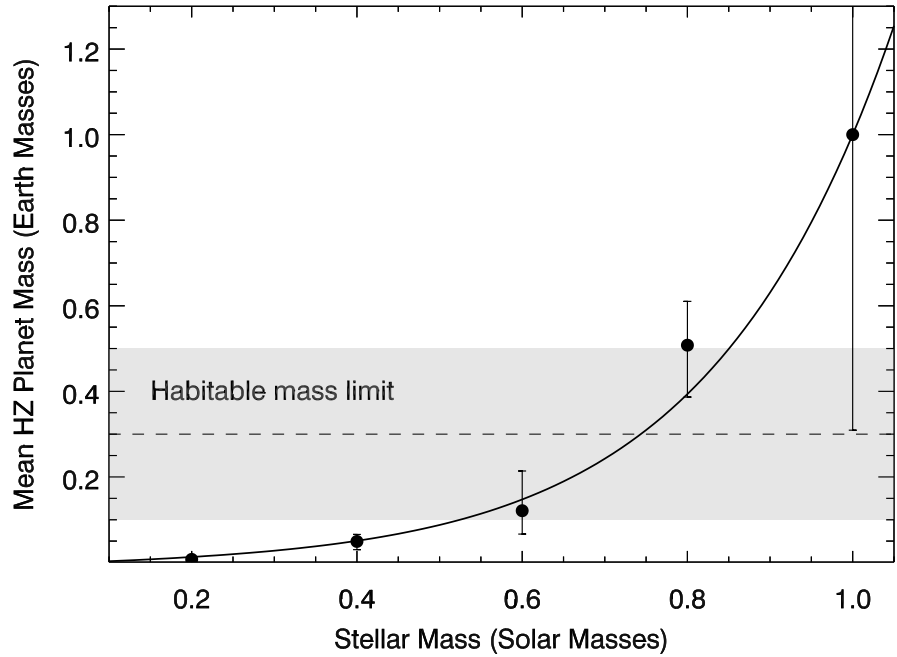}}
\caption{Mass of planets formed in the HZ as a function of stellar
mass, for a model with $h=1$, $\alpha=1$, and $fZ = 1.2$ (so that the mean
planet mass for a 1 $\msun$ star is 1 $\mearth$).  Error bars represent the
range of values for HZ planets.  The solid curve represents a model in which
the HZ planet mass scales linearly with the total annular mass in the HZ.  The
shaded region represents reasonable estimates of the limiting planet mass for
habitability (0.1-0.5 $\mearth$); our chosen value of 0.3 $\mearth$ is
indicated with the dashed line.}
\label{fig:mpl}
\end{figure}

\begin{figure}
\centerline{\epsscale{0.9}\plotone{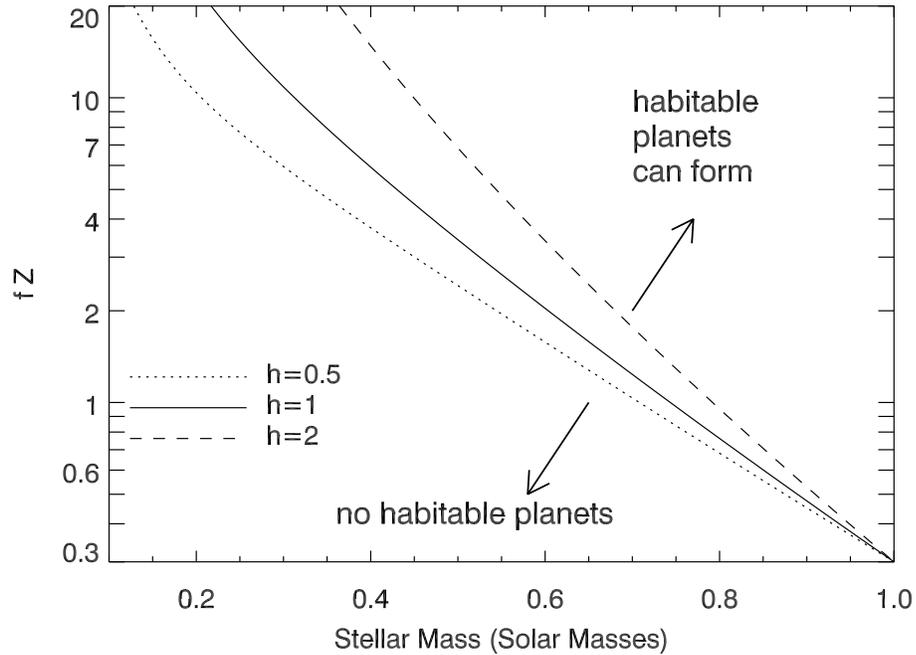}}
\caption{Regions of $M_\star-fZ$ space in which habitable planets more
massive that 0.3 $\mearth$ can form, assuming $\alpha=1$ and for three
dif{f}erent values of $h$.  Planets larger than 0.3 $\mearth$ can form above
and to the right of each curve.  }
\label{fig:mstar-h}
\end{figure}

\begin{figure}
\centerline{\epsscale{0.9}\plotone{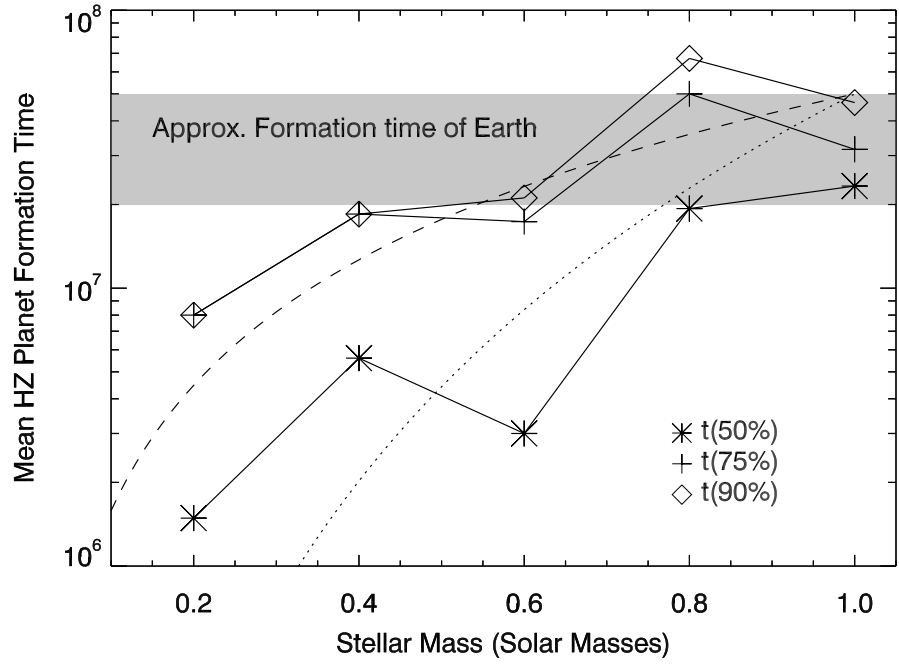}}
\caption{Formation times of HZ planets in our simulations.  Dif{f}erent symbols
correspond to the time for a planet to reach a fraction (50\%,75\%,90\%) of
its final mass.  Shaded are estimates for the formation time of the Earth,
derived from Hf/W isotopic measurements (e.g., Jacobsen 2005).  The dotted
line corresponds to a simple estimate from Safronov (1969), assuming the
formation time scales inversely with the product of the orbital frequency and
the local surface density.  The dashed line represents a dif{f}erent, simple
model in which the formation time scales inversely as the product of the
orbital {\it velocity} and the local surface density.  Both estimates are
referenced to 50 Myr for 1 $\msun$.}
\label{fig:tf}
\end{figure}

\begin{figure}
\centerline{\epsscale{1.0}\plotone{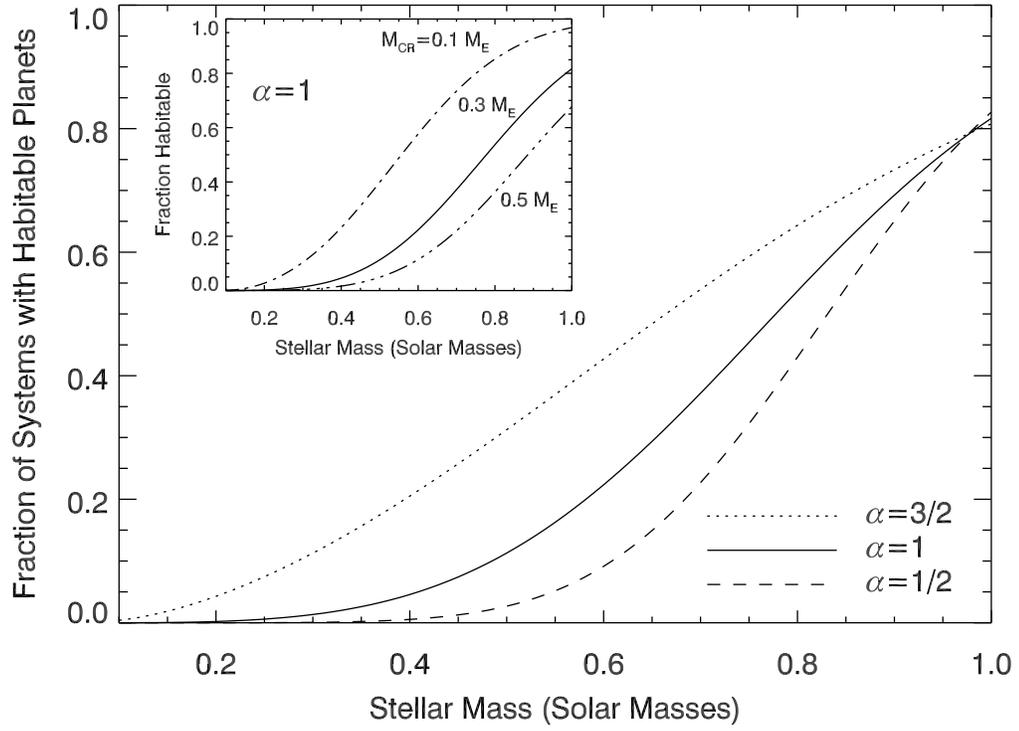}}
\caption{Fraction of stars able to form a $>0.3\mearth$ HZ planet as a
function of stellar mass.  We assume an $fZ$ distribution that is gaussian
with a standard deviation of 0.5 dex, and a fixed value of $h=1$.  The inset
shows the ef{f}ect of varying the limiting mass for habitability from 0.1 to
0.5 $\mearth$, for $\alpha=h=1$.}
\label{fig:frac}
\end{figure}

\end{document}